\documentclass[twocolumn]{aastex631}
\RequirePackage{graphicx}
\usepackage{acro}
\usepackage{color}

\DeclareAcronym{LIGO}{
  short = LIGO ,
  long = Laser Interferometer Gravitational-Wave Observatory ,
  short-plural = ,
}
\DeclareAcronym{GW}{
  short = GW ,
  long = gravitational wave ,
  short-plural = s ,
}
\DeclareAcronym{BBH}{
  short = BBH ,
  long = binary black hole ,
  short-plural = s ,
}
\DeclareAcronym{GR}{
  short = GR ,
  long = general relativity ,
  short-plural =  ,
}
\DeclareAcronym{PSD}{
  short = PSD ,
  long = power spectral density ,
  short-plural = s ,
}
\DeclareAcronym{PDF}{
  short = PDF ,
  long = probability distribution function ,
  short-plural = s ,
}
\DeclareAcronym{BF}{
  short = BF ,
  long = Bayes factor ,
  short-plural = s ,
}

\shorttitle{}
\shortauthors{Zhao et al.}
\graphicspath{{./}{figures/}}

\begin{document}

\title{Search for the birefringence of gravitational waves with the third observing run of Advanced LIGO-Virgo}

\author{Zhi-Chao Zhao}
\affiliation{Department of Astronomy, Beijing Normal University, Beijing 100875, P. R. China}

\author{Zhoujian Cao}
\affiliation{Department of Astronomy, Beijing Normal University, Beijing 100875, P. R. China}
\affiliation{School of Fundamental Physics and Mathematical Sciences, Hangzhou Institute for Advanced Study, University of Chinese Academy of Sciences, Hangzhou 310024, P. R. China}

\author{Sai Wang}
\affiliation{Theoretical Physics Division, Institute of High Energy Physics, Chinese Academy of Sciences, Beijing 100049, P. R. China}
\affiliation{School of Physical Sciences, University of Chinese Academy of Sciences, Beijing 100049, P. R. China}

\correspondingauthor{Sai Wang}
\email{wangsai@ihep.ac.cn}



\begin{abstract}

Gravitational waves would attain birefringence during their propagation from distant sources to the Earth, when the CPT symmetry is broken. 
If it was sizeable enough, such birefringence could be measured by the Advanced LIGO, Virgo and KAGRA detector network. 
In this work, we place constraints on the birefringence of gravitational waves with the third observing run of this network, i.e. two catalogues GWTC-2 and GWTC-3. 
For the dispersion relation $\omega^{2}=k^{2}\pm2\zeta k^{3}$, our analysis shows the up-to-date strictest limit on the CPT-violating parameter, i.e. \textcolor{black}{{$\zeta=4.07^{+5.91}_{-5.79}\times10^{-17}\mathrm{m}$}}, at $68\%$ confidence level. 
This limit is stricter by $\sim$5 times than the existing one
{($\sim 2\times10^{-16}$m)}
and stands for the first $\sim$10GeV-scale test of the CPT symmetry in gravitational waves. The results of Bayes factor strongly disfavor the birefringence scenario of gravitational waves. 

\end{abstract}



\section{Introduction} \label{sec:intro}

The CPT transformation is a fundamental symmetry in modern physics \cite{Peskin:1995ev}.
The results from numerous experiments in laboratories and astronomy have been shown to be consistent with the predictions of the CPT symmetry to high precision \cite{Kostelecky:2008ts,Will:2014kxa}, since it was first proposed in 1950s \cite{1951PhRv...82..914S}. 
Although no definitive signals of CPT violation have been uncovered, there are quantities of motivations, e.g. candidate theories of quantum gravity on Planck scale \cite{Kostelecky:1988zi,Kostelecky:1991ak,Amelino-Camelia:1997ieq,Amelino-Camelia:2008aez,Mielczarek:2017cdp}, to perform careful investigations on possible mechanisms and manifestations of CPT symmetry breaking. 
However, almost all tests of the CPT symmetry have been implemented in the electromagnetic or/and neutrino sectors, rather than the pure gravitational sector (see reviews \cite{Kostelecky:2008ts,Will:2014kxa} and references therein).

The discovery of \acp{GW} by the Advanced \ac{LIGO} \cite{LIGOScientific:2016aoc} in September 2015 opened a new observational window to testing the CPT symmetry, in particular, in the gravitational sector \cite{Kostelecky:2016kfm,Wang:2020pgu,Shao:2020shv,Wang:2021gqm,Wang:2020cub,Wang:2021ctl,Niu:2022yhr}. When the CPT symmetry is violated, \acp{GW} would attain birefringence during their propagation from distant sources to our detectors \cite{Kostelecky:2016kfm}. 
The birefringence could slightly widen
{or split }
the peak of the gravitational waveform \cite{Yamada:2020zvt}. 
The first constraint on the dimension-5 CPT-violating operators is reported to be smaller than $2\times10^{-14}\mathrm{m}$ \cite{Kostelecky:2016kfm}, by considering the width of peak at the maximal amplitude of the first event, i.e. GW150914. 
The birefringence could also lead to a rotation between the plus and cross modes of \acp{GW} \cite{Mewes:2019dhj,Wang:2017igw}.
In the spirit of effective field theory, the leading-order CPT-violating contribution to the gravitational waveform was evaluated quantitatively \cite{Mewes:2019dhj,Wang:2017igw}.
Based on this, we performed the first full Bayesian test of the CPT symmetry in Ref.~\cite{Wang:2020pgu}, by analyzing the events observed during the Advanced \ac{LIGO}-Virgo's first (O1) and second (O2) observing runs, i.e. GWTC-1 \cite{LIGOScientific:2018mvr}. 
No evidence of the CPT violation yielded a constraint on the CPT-violating parameter, i.e. $1.4^{+2.2}_{-3.1}\times10^{-16}\mathrm{m}$, which is two orders of magnitude stricter than before. 
Similar results were reported soon in Refs.~\cite{Wang:2020cub,Shao:2020shv}. 
Recently, Refs.~\cite{Wang:2021gqm,Wang:2021ctl} reported a new upper limit $4.5\times10^{-16}\rm{m}$ by using the third open gravitational-wave catalog \cite{Nitz:2021uxj}, which includes the events during the first half of the third observing run (O3a). {Model-dependent studies on the CPT symmetry have also been broadly investigated}\footnote{For example, see Refs.~\cite{Kostelecky:2011qz,Yagi:2017zhb,Yagi:2012vf,Crisostomi:2017ugk,Nishizawa:2018srh,Horava:2009uw,Gao:2019liu,Conroy:2019ibo,Alexander:2009tp,Jackiw:2003pm,Wu:2021ndf,Gong:2021jgg,Qiao:2021fwi,Takahashi:2009wc,Yoshida:2017cjl,Wang:2012fi,Zhu:2013fja,Wang:2017brl,Amelino-Camelia:2000cpa,Amelino-Camelia:2000stu,Kowalski-Glikman:2001vvk,Magueijo:2001cr,Gambini:1998it,Alfaro:2001rb,Carroll:2001ws,Douglas:2001ba,Kamada:2021kxi}}.
However, in either case, there has not been an analysis based on all three observing runs of Advanced \ac{LIGO}-Virgo. 

In this work, we revisit the Bayesian test of the CPT symmetry with the data in {GWTC-2, GWTC-2.1} \cite{LIGOScientific:2020ibl,LIGOScientific:2021usb} and GWTC-3 \cite{LIGOScientific:2021djp}.
The former includes the events observed during O3a, while the latter includes those during the second half of the third observing run (O3b). 
Besides a significant enlargement of the number of events, there are improvements on the performance of detectors and on the accuracy of template {approximates} \cite{LIGOScientific:2020ibl,LIGOScientific:2021usb,LIGOScientific:2021djp}. 
All of these possibly enable more accurate extraction of physical parameters from the observed events.
Therefore, we expect to perform a stricter test of the CPT symmetry in this paper. 

This paper is organized as follows. In Sec.~\ref{theory}, we review the gravitational waveform under the hypothesis of CPT violation and the method used for data analysis. In Sec.~\ref{result}, the results and discussions are shown explicitly. Finally, our conclusions can be found in Sec.~\ref{conclusions}. Throughout this paper we adopt $c=G=\hbar=1$, where $c$, $G$, and $\hbar$ denote the speed of light, Newtonian gravitational constant, and reduced Planck constant, respectively.

\section{Theory and Data analysis}
\label{theory}

The birefringence alters in an opposite way the phases of two circular polarization modes of \acp{GW}, though it contributes little to the amplitude of \acp{GW} \cite{Zhao:2019xmm}. 
For the leading-order CPT-violating effect, which is characterized by an independent parameter $\zeta$, the dispersion relation of \acp{GW} is given by 
\begin{equation}\label{eq:dispersionrelation}
    \omega^{2}=k^{2} \pm 2 \zeta k^{3}\ ,
\end{equation}
where the symbol $\pm$ stands for the birefringence, $\omega$ and $k$ denote the energy and momentum of \acp{GW}, respectively. 
We take ``$+$'' and ``$-$'' for the left- and right-handed polarization modes, respectively. 
The effects of higher-order CPT violation are neglected in this work, since they are expected to be suppressed by high energy scales \cite{Kostelecky:2003fs}. 

The \ac{GW} strain involving the birefringence is given as $h_{\mathrm{L}, \mathrm{R}}=h_{\mathrm{L}, \mathrm{R}}^{\mathrm{GR}} \mathrm{e}^{\pm i \delta \Psi}$ \cite{Mewes:2019dhj,Wang:2017igw}, where $h^{\rm GR}$ denotes the strain in \ac{GR}. 
For the left-/right-handed polarization mode, the phase is shifted by a factor, i.e. \cite{Mewes:2019dhj,Wang:2017igw}
\begin{equation}\label{eq:phasechange}
    \delta \Psi=4 \pi^{2} \zeta f^{2} \int_{0}^{z}\frac{ 1+z^\prime }{ H(z^\prime)} d z^\prime \ ,
\end{equation}
where $z$ is the redshift of the source, $f=\omega/2\pi$ is the frequency of \acp{GW} in the observer frame, and $H(z^\prime)$ is the Hubble parameter at redshift $z^\prime$.
Throughout this work, we use in our evaluation the cosmological parameters measured by Planck satellite 2015 \cite{Planck:2015fie}.

During the process of data analysis, the \acp{GW} are commonly decomposed in terms of the plus and cross modes, which are related to the left and right handed modes by following $h_{\mathrm{L}, \mathrm{R}}=h_{+} \pm i h_{\times}$. 
Therefore, the waveform involving the birefringence is given by \cite{Mewes:2019dhj,Wang:2017igw} 
\begin{equation}\label{template}
    \left(\begin{array}{c}
    h_{+} \\
    h_{\times}
    \end{array}\right)=\left(\begin{array}{cc}
    \cos (\delta \Psi) & -\sin (\delta \Psi) \\
    \sin (\delta \Psi) & \cos (\delta \Psi)
    \end{array}\right)\left(\begin{array}{l}
    h_{+}^{\mathrm{GR}} \\
    h_{\times}^{\mathrm{GR}}
    \end{array}\right)\ ,
\end{equation}
where $h_+^{\mathrm{GR}}$ and $h_\times^{\mathrm{GR}}$ stand for the plus and cross modes of the \ac{GR} waveform generated with the state-of-the-art ``IMRPhenomXPHM'' method \cite{Pratten:2020ceb}.
Based on Eq.~(\ref{eq:phasechange}) and Eq.~(\ref{template}), when the parameter $\zeta$ vanishes, the birefringence waveform would be recovered to the \ac{GR} waveform, as expected. 
In the following, we estimate the allowed value of $\zeta$ as well as the independent parameters of \ac{GR} waveform by performing data analysis. 

To infer the parameter space, we perform Bayesian analysis of the transient events in two recently released catalogues GWTC-2 \cite{LIGOScientific:2020ibl,LIGOScientific:2021usb} and GWTC-3 \cite{LIGOScientific:2021djp}. 
Specifically, we analyze the data of \acp{BBH} by using a modified version of  \texttt{pBilby} \cite{Smith:2019ucc} and \texttt{dynesty} \cite{skilling2004nested,Speagle:2019ivv}.
As discussed in Ref.~\cite{Wang:2020pgu}, the compact binary coalescence events involving neutron stars would be discarded in such an analysis, since they might be related to unknown matter effects rather than the pure-gravity effect. 
Therefore, the current work stands for an analysis of $65$ \ac{BBH} events in total. 
In addition, to check our inference configuration, we have reproduced the results of GWTC-2 and GWTC-3 without considering the birefringence effect.

{The log-likelihood function for a strain signal with Gaussian noise is defined as} \cite{Finn:1992wt,Thrane:2018qnx} 
\begin{equation}\label{bayesA}
\log\mathcal{L}(s |\vec{\xi},h)=\langle s,h(\vec{\xi}) \rangle-\frac{1}{2}\langle h(\vec{\xi}),h(\vec{\xi}) \rangle\ ,
\end{equation}
where $s$ denotes the strain signal, and $h(\vec{\xi})$ is the waveform template with independent parameters $\vec{\xi}$. 
For the birefringence waveform in Eq.~(\ref{template}), $\vec{\xi}$ also includes $\zeta$, besides the \ac{GR}-related parameters. 
In Eq.~(\ref{bayesA}), the inner product is defined as 
\begin{equation}
    \langle a,b \rangle=4\Re \int_{0}^{\infty}\frac{a(f)b^\ast(f)}{S_{n}(f)}df \ ,
\end{equation}
where $S_n(f)$ is the noise \ac{PSD} of a detector. 
In this work, the noise \ac{PSD} for each event, as well as the duration and minimum frequency cutoff configuration, is the same as that in either GWTC-2 or GWTC-3. 
To reduce the computational burden, the analytic marginalization procedures for the coalescence time and distance are used \cite{Ashton:2018jfp}. 
We assume the noise of multiple detectors to be uncorrelated, implying that the likelihoods of them can be multiplied.

Given a prior \ac{PDF} $p(\vec{\xi})$ and the likelihood function $\mathcal{L}(h|\vec{\xi})$, following Bayes' rule, we evaluate the posterior \ac{PDF} of $\vec{\xi}$, i.e. \cite{ramos2018bayesian}
\begin{equation}\label{BayesRule}
p(\vec{\xi} | s, h)=\frac{p(\vec{\xi}|h) \mathcal{L}(s | \vec{\xi},h)}{\mathcal{Z}(s|h)} \ ,
\end{equation}
{where the Bayesian evidence is defined as} \cite{Kass:1995loi} 
\begin{equation}
    \mathcal{Z}(s|h)=\int \mathcal{L}(s | \vec{\xi},h) p(\vec{\xi}|h) d \vec{\xi}\ .
\end{equation} 
In our parameter inference, the priors for the \ac{GR}-related parameters are the same as those used in GWTC-2 and GWTC-3. 
For the prior of $\zeta$, we introduce a uniform distribution over $[-4,4]$ in units of $10^{-14}$ metres.\footnote{{For GW191204\_110529, we employ a uniform distribution over $[-40,40]$ instead of $[-4,4]$, since the latter would produce a posterior that touches the boundaries of the prior}.}

To perform model comparison, we employ the \ac{BF} which is a ratio between the Bayesian evidence of \ac{GR} and birefringence models, i.e.  \cite{ramos2018bayesian}
\begin{equation}
{\rm{BF}} = \frac{\mathcal{Z}(s|h^{\rm{GR}}) }{\mathcal{Z}(s|h)}\ ,
\end{equation}
where $h^{\rm{GR}}$ and $h$ denote the waveforms in \ac{GR} and the birefringence scenario, respectively. 
For multiple events, the total \ac{BF} is obtained by multiplying the \acp{BF} of them together.
The value of \ac{BF} shows which model is more favored by the data \cite{Reyes:2019bfv,Trotta:2008qt}.

\section{Result and Discussion}
\label{result}

\begin{table*}[htb]
\caption{{Median value and $90\%$ confidence intervals} of the parameter $\zeta$ for each event in GWTC-2 (left columns) and GWTC-3 (right columns). Bayes factor is shown to compare \ac{GR} and the birefringence scenario. }
\label{tab:1}
\begin{tabular}{lrr|lrr}
  \hline
{~~~~~~~~~Event} &  $\zeta ~[ 10^{-14} {\rm m}]$          & BF \quad&\quad {~~~~~~~~~Event} &  $\zeta ~[ 10^{-14} {\rm m}]$          & BF \\
\hline
GW190408\_181802 &  $-0.00^{+0.25}_{-0.25}$ &        17.47  & GW191103\_012549                     &  $-0.01^{+0.25}_{-0.30}$ &        35.11 \\
GW190412        &   $0.07^{+0.16}_{-0.25}$ &        16.99  & GW191105\_143521                     &   $0.00^{+0.37}_{-0.39}$ &        38.59 \\
GW190413\_052954 &   $0.09^{+0.27}_{-0.35}$ &         8.74  & GW191109\_010717                     &  $-0.16^{+0.46}_{-0.15}$ &         0.69 \\
GW190413\_134308 &  $-0.01^{+0.70}_{-0.71}$ &         4.67  & GW191113\_071753                     &  $-0.61^{+2.19}_{-0.99}$ &         1.44 \\
GW190421\_213856 &  $-0.08^{+0.78}_{-0.71}$ &         5.35  & GW191126\_115259                     &   $0.00^{+0.72}_{-0.75}$ &        42.29 \\
GW190424\_180648 &   $0.03^{+0.20}_{-0.25}$ &         6.23  & GW191127\_050227                     &   $0.03^{+0.79}_{-0.77}$ &         4.93 \\
GW190503\_185404 &  $-0.55^{+0.87}_{-0.44}$ &         3.50  & GW191129\_134029                     &   $0.01^{+0.04}_{-0.05}$ &        18.56 \\
GW190512\_180714 &   $0.01^{+0.12}_{-0.15}$ &        24.49  & GW191204\_110529                     &   $0.05^{+4.95}_{-4.80}$ &         3.52 \\
GW190513\_205428 &  $-0.06^{+0.28}_{-0.21}$ &        13.85  & GW191204\_171526                     &   $0.00^{+0.01}_{-0.01}$ &       320.34 \\
GW190514\_065416 &   $0.03^{+0.62}_{-0.66}$ &         6.05  & GW191215\_223052                     &   $0.02^{+0.09}_{-0.13}$ &        21.67 \\
GW190517\_055101 &   $0.14^{+0.24}_{-0.41}$ &        15.77  & GW191216\_213338                     &   $0.00^{+0.02}_{-0.02}$ &       348.53 \\
GW190519\_153544 &   $0.02^{+0.59}_{-0.62}$ &         8.87  & GW191222\_033537                     &   $0.02^{+0.77}_{-0.82}$ &         4.34 \\
GW190521        &   $0.22^{+1.05}_{-2.79}$ &         0.60  & GW191230\_180458                     &  $-0.04^{+0.60}_{-0.57}$ &         7.90 \\
GW190521\_074359 &   $0.03^{+0.59}_{-0.62}$ &         4.27  & GW200112\_155838                     &   $0.00^{+0.15}_{-0.11}$ &        33.65 \\
GW190527\_092055 &   $0.02^{+0.46}_{-0.56}$ &        14.43  & GW200128\_022011                     &  $-0.00^{+0.26}_{-0.25}$ &        12.79 \\
GW190602\_175927 &   $0.01^{+0.79}_{-0.84}$ &         4.73  & GW200129\_065458                     &  $-0.05^{+0.08}_{-0.06}$ &        28.21 \\
GW190620\_030421 &   $0.09^{+0.66}_{-0.76}$ &         5.89  & GW200202\_154313                     &  $-0.01^{+0.13}_{-0.22}$ &       142.85 \\
GW190630\_185205 &  $-0.01^{+0.34}_{-0.36}$ &        17.57  & GW200208\_130117                     &  $-0.07^{+0.16}_{-0.21}$ &        25.14 \\
GW190701\_203306 &   $0.22^{+0.49}_{-0.38}$ &         8.36  & GW200208\_222617                     &   $0.33^{+0.60}_{-1.25}$ &         1.02 \\
GW190706\_222641 &   $0.14^{+0.69}_{-0.87}$ &         3.62  & GW200209\_085452                     &   $0.00^{+0.23}_{-0.24}$ &        27.21 \\
GW190708\_232457 &  $-0.00^{+0.10}_{-0.11}$ &        42.56  & GW200210\_092255                     &  $-0.05^{+2.87}_{-2.78}$ &        11.79 \\
GW190719\_215514 &  $-0.00^{+1.46}_{-1.33}$ &         9.93  & GW200216\_220804                     &   $0.19^{+0.77}_{-0.79}$ &         5.51 \\
GW190727\_060333 &  $-0.01^{+0.29}_{-0.26}$ &        13.91  & GW200219\_094415                     &   $0.09^{+0.46}_{-0.60}$ &         5.41 \\
GW190731\_140936 &   $0.00^{+0.40}_{-0.40}$ &        13.17  & GW200220\_061928                     &   $0.15^{+1.55}_{-1.78}$ &         2.75 \\
GW190803\_022701 &   $0.03^{+0.41}_{-0.42}$ &        14.45  & GW200220\_124850                     &  $-0.02^{+0.46}_{-0.43}$ &        10.73 \\
GW190828\_063405 &  $-0.00^{+0.31}_{-0.36}$ &        14.21  & GW200224\_222234                     &   $0.01^{+0.22}_{-0.11}$ &        28.68 \\
GW190828\_065509 &  $-0.28^{+0.69}_{-0.16}$ &         1.34  & GW200225\_060421                     &   $0.01^{+0.09}_{-0.10}$ &        11.67 \\
GW190909\_114149 &  $-0.04^{+1.30}_{-1.21}$ &         2.10  & GW200302\_015811                     &   $0.00^{+0.15}_{-0.14}$ &        41.93 \\
GW190910\_112807 &   $0.00^{+0.38}_{-0.36}$ &        13.85  & GW200306\_093714                     &   $0.01^{+0.87}_{-0.95}$ &        11.15 \\
GW190924\_021846 &   $0.72^{+0.10}_{-1.50}$ &         1.14  & GW200308\_173609                     &  $-0.09^{+3.51}_{-3.36}$ &         1.11 \\
GW190929\_012149 &   $0.03^{+1.72}_{-1.60}$ &         1.93  & GW200311\_115853                     &  $-0.02^{+0.10}_{-0.08}$ &        34.94 \\
GW190930\_133541 &  $-0.00^{+0.14}_{-0.14}$ &        28.02  & GW200316\_215756                     &   $0.25^{+1.32}_{-1.77}$ &         7.65 \\
 ~~~~~~~~~~~& ~~~~~~~~~~~~& ~~~~~~~~~~~ & GW200322\_091133 &  $-0.08^{+3.38}_{-3.28}$ &         0.90 \\
  \hline
\end{tabular}
\end{table*}

\begin{figure}[ht!]
\plotone{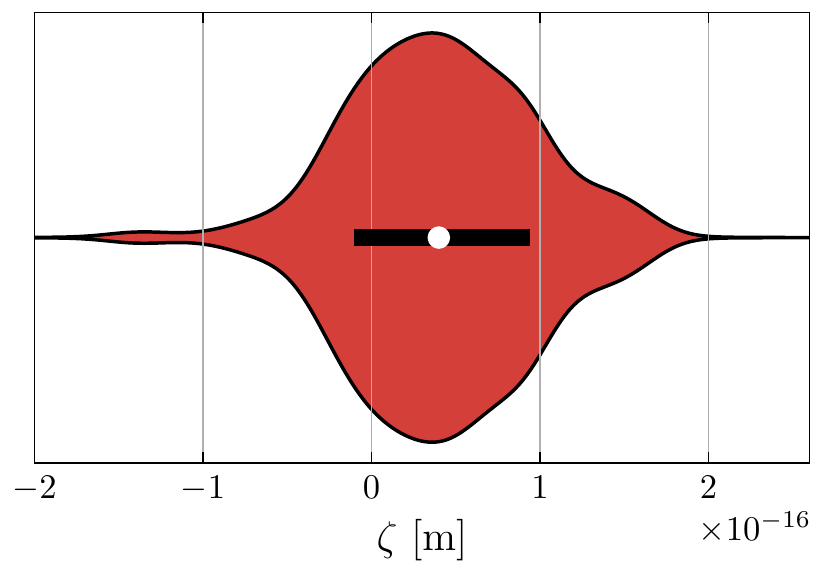}
\caption{Posterior \ac{PDF} of the parameter $\zeta$ from the joint analysis of all events. The median value is represented with 
a white point while the {$68\%$ confidence interval} with a black solid line.  \label{fig:JointConstraint}}
\end{figure}

The results of this work are shown in Table~\ref{tab:1} and Figure~\ref{fig:JointConstraint}. 
We show in Tab.~\ref{tab:1} the median value and {90\%} confidence interval of $\zeta$ for each event in GWTC-2 (left columns) and GWTC-3 (right columns). 
We also show the results of Bayes factor to compare \ac{GR} and the birefringence scenario. 
In Fig.~\ref{fig:JointConstraint}, we depict the posterior \ac{PDF} of $\zeta$ from a joint analysis of all events. 

Based on Tab.~\ref{tab:1}, we find that each event in GWTC-2 and GWTC-3 is well compatible with null birefringence, i.e. $\zeta=0$. 
There is not significant evidence for the CPT violation and birefringence in \acp{GW}. 
However, we obtain the up-to-date best constraints on $\zeta$. 
Amongst all events, GW191204\_171526 and GW191216\_213338 in GWTC-3 reveal the strictest bounds on $\zeta$, i.e. $\zeta= 0.8^{+6.4}_{-6.8} \times 10^{-17} {\rm m}$ and $\zeta= 2.3^{+6.6}_{-7.6} \times 10^{-17} {\rm m}$ {at $68\%$ confidence level}, respectively. 
These bounds are stricter by $\sim$3 orders of magnitude than the first upper bound $2\times10^{-14}\mathrm{m}$ \cite{Kostelecky:2016kfm}, which was obtained from the first event GW150914. 
They are also stricter by $\sim$1 order of magnitude than the upper bound $|\zeta| < \rm{few}\times 10^{-16} {\rm m}$ \cite{Wang:2020pgu,Shao:2020shv}, which was obtained from GWTC-1. 
Moreover, they are stricter than the upper bound $\sim4.5\times10^{-16}\mathrm{m}$ \cite{Wang:2021gqm,Wang:2021ctl}, which were obtained via a joint analysis of GWTC-1 and GWTC-2. 
Interestingly, we find that the aforementioned two events alone can lead to better bounds on $\zeta$ than those events in GWTC-1 and GWTC-2.

Combining the posterior \acp{PDF} of all events together, we can obtain a joint constraint on $\zeta$, which is stricter than the bounds from the individual events mentioned above. 
By employing \texttt{Monte Python} \cite{Audren:2012wb}, we obtain the combined bound to be 
\begin{equation}\label{eq:limit}
{\color{black}\zeta = 4.07^{+5.91}_{-5.79}\times10^{-17} \mathrm{m}}
\end{equation}
{at $68\%$ confidence level}. 
It stands for an upper bound of $|\zeta|<\rm{few}\times10^{-17}\rm{m}$, which is roughly corresponded to an energy scale of $\sim$10GeV. 
{{It becomes $4.1^{+12.4}_{-12.2}\times10^{-17} \rm{m}$ and $4.1^{+19.2}_{-20.1}\times10^{-17} \rm{m}$ at $95\%$ and $99.7\%$ confidence levels, respectively. Or equivalently, it is $\zeta = 4.07^{+10.2}_{-10.4}\times10^{-17}$ at $90\%$ confidence level.}}
Based on this joint analysis, we show a violin depiction for the posterior \ac{PDF} of $\zeta$ in Fig.~\ref{fig:JointConstraint}. We represent the median value with a white point while {the $68\%$ confidence interval} with a black solid line. 
The limit in Eq.~(\ref{eq:limit}) is not only better than the existing \ac{GW}-only bounds, as demonstrated in the previous paragraph, but also better than the electromagnetic bounds. To be specific, it is stricter by $\sim$23 orders of magnitude than the limit from the LAGEOS satellite in the Solar system \cite{Smith:2007jm}, by $\sim$19 orders of magnitude than the limit from double binary pulsar, i.e. PSR J0737-3039 A/B \cite{Ali-Haimoud:2011wpu}, and by $\sim$9 orders of magnitude than the limit obtained by measuring the propagation speed of \acp{GW} from the binary neutron star merger event, i.e. GW170817/GRB 170817A \cite{Nishizawa:2018srh}. 
Therefore, the limit in Eq.~(\ref{eq:limit}) stands for the up-to-date strictest constraint on $\zeta$.
Meanwhile, our analysis is the first search for the CPT violation and birefringence in \acp{GW} on the $\sim$10GeV scale. 

With combination of all the events analyzed in this work, the total \ac{BF} indicates very strong evidence for \ac{GR} rather than the birefringence scenario. 
To be specific, we find 3 events, i.e. GW191204\_171526, GW191216\_213338 and GW200202\_154313, to have the largest \acp{BF}, i.e. $320$, $349$ and $143$, respectively. 
These events indicate strong evidence for \ac{GR}, since they have ${\rm BF} \in [45,740]$. 
It is interesting to find that the strictest bounds on $\zeta$ also arise from the first two events. 
However, we do not find any special ingredients, e.g. the mass ratio or the higher harmonic modes, to account for why these two events have such large \acp{BF} than others. 
Other events do not show strong evidence for either scenario. 
However, there are 59 events having \acp{BF} greater than one. 
Amongst them, 41 events with ${\rm BF} \in [5.5,45]$ show moderate evidence for \ac{GR}, while 18 events with ${\rm BF} \in [1,5.5]$ indicate weak evidence. 
Exceptions include 3 events, i.e. GW190521, GW191109\_010717 and GW200322\_091133.
We show their \acp{BF} to be smaller than one. 
Due to ${\rm BF} \in [0.18,1.0]$, they just indicate weak evidence for the birefringence model.

\section{Conclusion}
\label{conclusions}

In this work, we constrained the CPT violation and birefringence in \acp{GW} by performing Bayesian analysis of GWTC-2 and GWTC-3. 
We reported no significant evidence for the deviations from \ac{GR}. 
The limits on the CPT-violating parameter $\zeta$ were shown to be  $\rm{few}\times10^{-17}\rm{m}$, corresponding to $\sim$10GeV energy scale. 
These are the up-to-date strictest limits from individual events. 
Meanwhile, we obtained the combined constraint on $\zeta$ by performing the joint analysis of GWTC-2 and GWTC-3. 
The result was shown to be \textcolor{black}{ $\zeta=4.07^{+5.91}_{-5.79}\times10^{-17}\mathrm{m}$ at $68\%$ confidence level, {or equivalently, $\zeta = 4.07^{+10.2}_{-10.4}\times10^{-17}$ at $90\%$ confidence level}}, which is obviously stricter than those of individual events. 
We found that the above limits are tighter than all the existing ones in the literature. 
On the other hand, we found 3 events to indicate weak evidence for the birefringence model, while the other 62 events to indicate strong or moderate or weak evidence for \ac{GR}. 
The total \ac{BF} was shown to indicate very strong evidence for \ac{GR} rather than the birefringence scenario. 
In this work, we have ignored possible unknown effects of CPT violation on the generation of \acp{GW} in the sources. 
However, these effects may be interesting and would be studied in the future. 
In addition, our method is suitable to analyze the data from upcoming Advanced \ac{LIGO} observing runs \cite{KAGRA:2013rdx}. 

\begin{acknowledgements}
  Z.C.Z. and Z.C. are supported by the National Natural Science Foundation of China (Grant No. 12005016, No. 11690023 and No. 12021003).
  S.W. is partially supported by the grants from the National Natural Science Foundation of China with Grant NO. 12175243, the Institute of High Energy Physics with Grant NO. Y954040101, and the Key Research Program of the Chinese Academy of Sciences with Grant NO. XDPB15. 
  {This research has made use of Parallel Bilby v1.0.1 \cite{pbilby_paper, bilby_paper}, a parallelised Bayesian inference Python package, and Dynesty v1.0.1 \cite{dynesty_paper, skilling2004, skilling2006}, a nested sampler, to perform Bayesian parameter estimation.}
\end{acknowledgements}


\bibliography{Main}{}
\bibliographystyle{aasjournal}


\end{document}